\documentclass{jfm}

\usepackage{graphicx}
\usepackage{natbib}

\title[Kelvin wake pattern at large Froude numbers]{Kelvin wake pattern at large Froude numbers}

\author[A. Darmon, M. Benzaquen and E. Rapha\"el]%
{Alexandre Darmon$^{1,2}$\thanks{These authors contributed equally to this work}, 
Michael Benzaquen$^1$$\dagger$\thanks{Email address for correspondence: michael.benzaquen@espci.fr}  and Elie Rapha\"el$^1$}

\affiliation{$^1$PCT, UMR CNRS 7083 Gulliver, ESPCI ParisTech, 10 rue Vauquelin, 75005 Paris, France\\[\affilskip]
$^2$EC2M, UMR CNRS 7083 Gulliver, ESPCI ParisTech, 10 rue Vauquelin, 75005 Paris, France}

\begin{document}

\maketitle

\begin{abstract}
Gravity waves generated by an object moving  at constant speed at the water surface form a specific pattern commonly known as the Kelvin wake. It was proved by Lord Kelvin that such a wake is delimited by a constant angle $\simeq 19.47^\circ$. However a recent study by Rabaud and Moisy based on the observation of airborne images showed that the wake angle seems to decrease as the Froude number $Fr$ increases, scaling as $Fr^{-1}$ for large Froude numbers. To explain such observations the authors make the strong hypothesis that an object of size $b$ cannot generate wavelengths larger than $b$.  With no need of such an assumption and   modelling the moving object by an axisymmetric pressure field, we analytically show that the angle corresponding to the maximum amplitude of the waves scales as $Fr^{-1}$  for large Froude numbers, whereas the angle delimiting the wake region outside which the surface is essentially flat remains constant and equal to the Kelvin angle for all $Fr$.
\end{abstract}


\section{Introduction}
Simply by looking at a duck swimming in a pond or a cargo ship moving on a calm sea, one can clearly tell that there is something common about their wake. Indeed, they both display a familiar V-shaped pattern which only differ from one another by their dimensions. In 1887, Lord Kelvin \citep{Kelvin1887} proved that the wake created by an object moving at a uniform pace is always delimited by an angle equal to $\arcsin 1/3 = 19.47^\circ$. This theory, based on stationary phase arguments, is widely used both at theoretical and technical levels \citep{Lighthill1978,Lamb1993,Darrigol2005,Parnell2001}. Since Lord Kelvin, other studies have shown that two sets of waves can be distinguished in the wake: the so-called transverse waves and diverging waves \citep{Crawford1984,Nakos1990}. Their amplitudes directly depend on the hull Froude number  $Fr = V/\sqrt{gb}$, where $V$ is the speed of the moving object and $b$ its typical size. It has been observed that as the Froude number increases, so does the amplitude of the diverging waves; but that of the transverse waves decreases rapidly  and seems to vanish for sufficiently high Froude numbers \citep{Lighthill1978}. Speedboats, whose range of reachable Froude number is large, typically until $Fr \simeq 3$, experience different regimes as the Froude number is increased,  eventually entering the so-called planing regime in which their drag is significantly decreased as they displace fewer quantities of water  \citep{Cumbertach1958, Casling1978, Lai1995}. The understanding of the wave drag is of great practical importance in the ship industry for the hull design \citep{Tuck2002,Suzuki1997,Rabaud2013Autre}. At smaller scales, the wave drag for capillary gravity waves has also been studied extensively \citep{LeMerrer2011,Benzaquen2011,Benzaquen2012} and is notably of interest for insect locomotion \citep{Voise2010,Chepelianskii2008}.
\smallskip

Recent experimental observations by Rabaud and Moisy have challenged the commonly accepted theory of Lord Kelvin \citep{rabaud2013ship}. Based on aiborne observations of ship wakes, the authors show that the wake angle seems to decrease as the Froude number is increased, scaling as $Fr^{-1}$ for large Froude numbers. To explain their observations, they make the strong hypothesis that an object of size $b$ cannot generate wavelengths greater than $b$. Even though this assumption leads to consistent results   regarding the experimental findings, it has not been firmly established and is open to questioning.
\smallskip

We here propose an explanation of such observations with no need of the above mentioned maximum wavelength argument. We first perform numerical evaluation of the surface displacement induced by a moving pressure field of typical size $b$ above the surface of water and show that two angles can actually be highlighted in the wake: the outer angle delimiting the wake, shown to be constant and equal to the Kelvin angle, and an inner angle corresponding to the maximum amplitude of the waves.  We then analytically  prove that the latter is not constant and scales as $Fr^{-1}$ at large Froude numbers.
\smallskip

\section{Surface displacement}\label{sec:surface_displacement}

In the pure gravity waves limit, the surface displacement generated by a pressure field $p(x,y)$ moving in the $-x$ direction with constant speed $V$ can be written in the frame of reference of the moving perturbation as \citep{Havelock1908,Havelock1919,Raphael1996}:
\begin{eqnarray}
\zeta(x,y) = -\lim_{\varepsilon\rightarrow 0}   \, \int \!\!\! \int \frac{\mathrm{d} k \, \mathrm{d} \theta}{4\pi^2\rho}  \, \frac{\hat{p}(k,\theta) \, e^{-ik(\cos \theta \, x-\sin \theta \, y)}}{c(k)^2-V^2 \cos^2 \theta+   2 i \varepsilon V \cos\theta /k} \, , 
\label{profil12}
\end{eqnarray}
where  $\hat{p}(k,\theta)$ is the Fourier Transform of $p(x,y)$ in cylindrical coordinates, $\rho$ is the water density and $c(k)=(g/k)^{1/2}$ is the phase speed for pure gravity waves.
Let us now nondimensionalise the problem through:
\begin{eqnarray}
\displaystyle Z=\frac{ 4 \pi^2\zeta }b  \, ; \,    \,  X= \frac x b  \, ; \,\, Y=\frac y b\, ; \, \,K= k b \,; \, \,
\displaystyle\widehat{P}=\frac{\hat{p}}{\rho g b^3} \,;  \,\, \tilde{\varepsilon}=\frac{\varepsilon}{\sqrt{g/b}}     \, , 
\label{ND}
\end{eqnarray} 
where $b$ is the typical size of the pressure field $p(x,y)$. Equation (\ref{profil12}) together with Eq. (\ref{ND})  becomes:
\begin{eqnarray}
Z(X,Y)&=&\int_{-\pi/2}^{\pi/2}  \mathrm{d} \theta \, F(\theta,X,Y)        \, ,\label{profiladim}
\end{eqnarray}
where:
\begin{eqnarray}
F(\theta,X,Y) = -\lim_{\tilde{\varepsilon}\rightarrow 0}\, \int_0^\infty K\, \mathrm{d} K \, \frac{\widehat{P}(K,\theta) \, e^{-iK\, (\cos \theta \, X-\sin \theta \, Y)}}{1-Fr^2\,K\, \cos^2 \theta + 2i\,\tilde{\varepsilon} \, Fr\,\cos \theta} \ . 
\label{F}
\end{eqnarray}
Using the Sokhotski-Plemelj formula (see \textit{e.g.} \citep{Appel2007}), one can write: 
\begin{eqnarray}
          F(\theta,X,Y) &=&  i \pi \Phi(K_0,\theta,X,Y)+G(\theta,X,Y) \, , \label{plejmel}
\end{eqnarray}
where:
\begin{eqnarray}
\Phi(K,\theta,X,Y)&=&\frac{K\, \widehat{P}(K,\theta) \, e^{-iK\, (\cos \theta \, X-\sin \theta \, Y)}}{Fr^2\cos^2 \theta} \ ,  \label{Phi}\\
K_0(\theta)&=&\frac{1}{Fr^2\cos^2 \theta} \ , \label{K0}
          \end{eqnarray}
and $\int  \mathrm{d} \theta\, G(\theta,X,Y)$ is a rapidly decreasing function with the distance to the perturbation.
According to Eqs. (\ref{profiladim}), (\ref{plejmel}), (\ref{Phi}) and (\ref{K0}) and sufficiently far from the perturbation, the surface displacement is well approximated by: 
\begin{eqnarray}
Z(X,Y)\simeq i\pi \int_{-\pi/2}^{\pi/2}  \mathrm{d} \theta \,    \frac{ \widehat{P}(K_0(\theta),\theta) \, e^{-i {(\cos \theta \, X-\sin \theta \, Y)}/{(Fr^2\cos^2\theta)} }}{Fr^4\cos^4 \theta}    \, .\label{Numint}
\end{eqnarray}
For a given pressure distribution one can thus obtain the surface displacement by numerically evaluating the integral in Eq. (\ref{Numint}).

\begin{figure}
  \centering \includegraphics[scale=0.33]{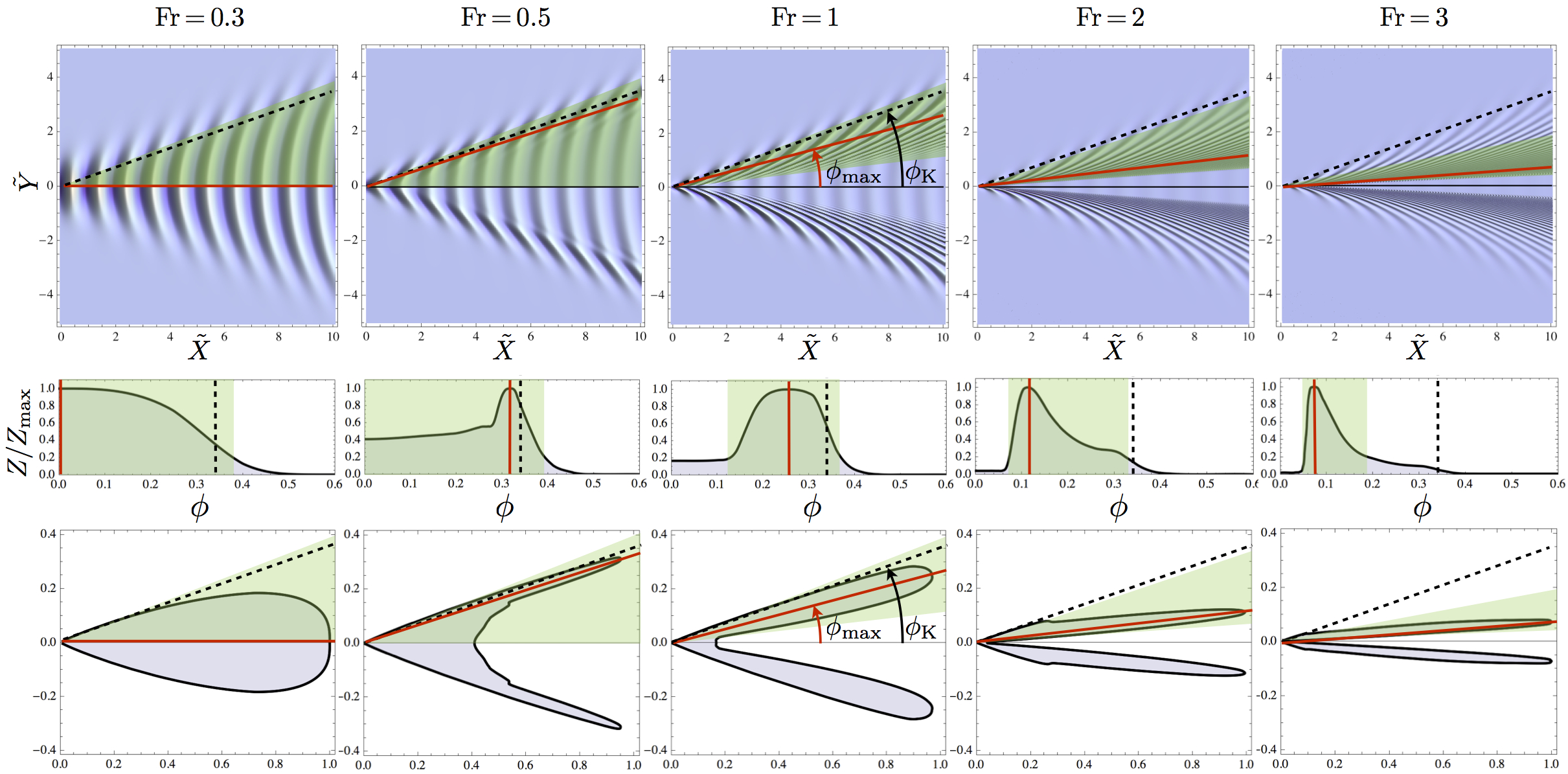}
 \caption{Colour online. 1st row: Relief plots of the surface displacement computed using Eq. (\ref{Numint}) for different Froude numbers as a function of $\tilde X=X/\Lambda$  and $\tilde Y=Y/\Lambda$ where $\Lambda=2\pi Fr^2$ is the dimensionless wavelength. 2nd row: Plot of the normalised angular envelope of the surface displacement as a function of $\phi$ for different Froude numbers where $\phi$ is the polar angle originating at the horizontal axis. The angular envelope was calculated by interpolating the maxima of the waves amplitudes over one wavelength sufficiently far from the perturbation. 3rd row: Polar plot of the normalised angular envelope as defined in the 2nd row for different Froude numbers. In all graphs the maximum of the angular envelope obtained for $\phi=\phi_{{max}}$ is signified with a solid red line and the Kelvin angle $\phi_ K={\arcsin(1/3)}$ is represented by a dashed black line. The green coloured region delimits the area in which the amplitude of the waves is above 20$\%$ of the maximum of the angular envelope.}
\label{profils}
\end{figure}

\section{Numerical evaluation}\label{sec:numerical_evaluation}
Considering a Gaussian pressure field of typical size $b$, symmetrical around the origin, with corresponding Fourier transform of the form: 
\begin{eqnarray}\hat P(K) = \exp[{-K^2/(4\pi^2)} ] \ , \label{Gauss}
\end{eqnarray}
Eq. (\ref{Numint}) yields the profiles displayed in Fig.~\ref{profils}. The first row shows  relief plots of the surface displacement computed using Eq.~(\ref{Numint}) for different Froude numbers as a function of $\tilde X=X/\Lambda$  and $\tilde Y=Y/\Lambda$ where $\Lambda=2\pi Fr^2$ is the dimensionless wavelength. The second row displays the normalised angular envelope of the surface displacement as a function of $\phi$ for different Froude numbers where $\phi$ is the polar angle originating at the horizontal axis. The angular envelope was calculated by interpolating the maxima of the waves amplitudes over one wavelength sufficiently far from the perturbation. The third row displays polar plots of the normalised angular envelope as defined in the second row for different Froude numbers. In all graphs the maximum of the angular envelope obtained for $\phi=\phi_{{max}}$ is signified with a solid red line and the Kelvin angle $\phi_ K={\arcsin(1/3)}$ is represented by a dashed black line. The green coloured region delimits the area in which the amplitude of the waves is above 20$\%$ of the maximum of the angular envelope. We chose this as an arbitrary criteria for what the eye can see. It somehow sets the error bars to the angle determined by just looking at a boat's wake on an airborne picture.

\begin{figure}
\begin{center}
\includegraphics[scale=0.363]{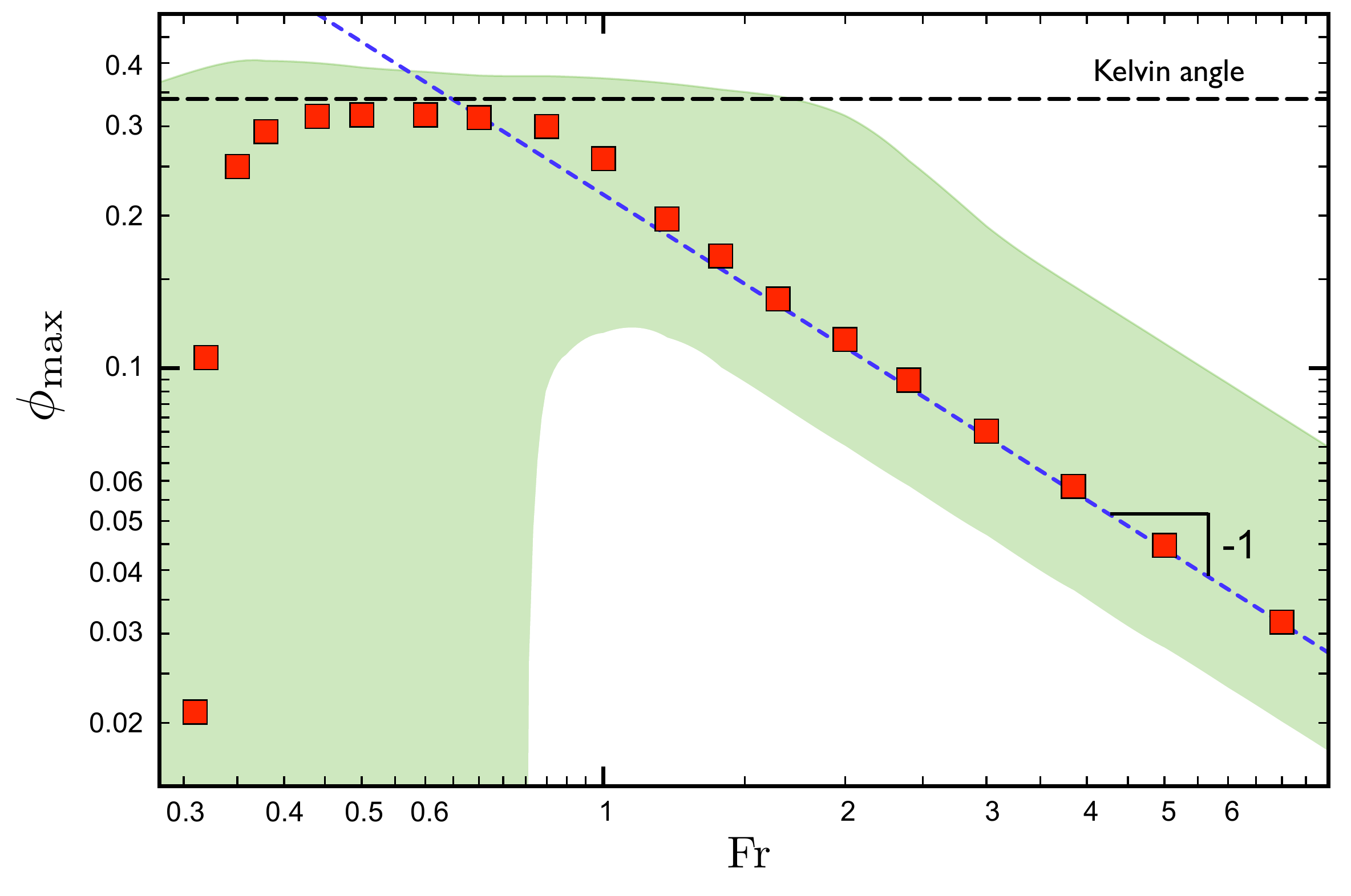}
\caption{Colour online. Plot of $\phi_{{max}}$ as defined in paragraph 2. as a function of the Froude number (red squares). The green coloured region delimits the area in which the amplitude of the waves is above 20$\%$ of the maximum of the angular envelope as defined in paragraph 2. The dashed black line signifies the Kelvin angle $\phi_ K={\arcsin(1/3)}$. The dashed blue line represents the asymptotic theoretical prediction at large Froude numbers as given by Eq. (\ref{scaling}).}
\label{Beta}
\end{center}
\end{figure} 

Firstly, as one can see, the wake pattern strongly depends on the Froude number. At $Fr=0.5$  one can clearly distinguish two different sets of waves: the so-called transverse waves that are orthogonal to the trajectory, and the so called diverging waves that are located at the edges of the wake. As the Froude number is increased the relative amplitude of the transverse waves decreases until vanishing making way for the diverging waves.  Secondly, we are interested in the evolution of $\phi_{{max}}$ as the Froude number increases.  Fig. \ref{Beta} displays
$\phi_{{max}}$ as a function of the Froude number (red squares). The green coloured region has the same meaning as that of Fig.~\ref{profils}. The dashed black line signifies the Kelvin angle $\phi_ K={\arcsin(1/3)}$. The dashed blue line represents the asymptotic scaling $Fr^{-1}$. At $Fr=0.3$ and below, the maximum of the wake's angular envelope is located at the central line $\phi=0$. A continuous transition in which the maximum is displaced to the edges of the wake  $\phi_{{max}}\simeq \phi_{{K}}$  is observed at $Fr\simeq0.31$. The angle $\phi_{{max}}$ then remains constant until $Fr\simeq0.7$ before it starts to decrease eventually scaling as $Fr^{-1}$. As one can see in Fig.~\ref{Beta} the green beam also scales as $Fr^{-1}$ for large Froude numbers thus explaining the observations of Rabaud and Moisy \citep{rabaud2013ship}. Thirdly, the waves are always confined within the Kelvin wake and always reach its outer boundary (see Fig.~\ref{profils}), even though the relatively small amplitude around this region can make it difficult to see on photographs as it might be diluted in the noise of the open sea. For a clear photograph where both the maximum amplitude angle and the Kelvin angle can be clearly identified see p. 96 of \citep{Falkovich2011}.

\section{High Froude numbers}\label{sec:high_froude_numbers}

In the following we demonstrate analytically the $\phi_{{max}}\sim Fr^{-1}$ scaling for large Froude numbers. Surface displacement as given by Eq. (\ref{Numint}) can be expressed in polar coordinates $X=R\cos \phi$, $Y= R\sin \phi$ as:
\begin{eqnarray}
\breve Z(R,\phi)\simeq i\pi \int_{-\pi}^{\pi}  \mathrm{d} \theta \,    \frac{ \widehat{P}(K_0(\theta),\theta)  e^{-iR \cos (\theta +\phi) /(Fr^2\cos^2\theta) }}{Fr^4\cos^4 \theta}    \, .\label{NumintPol}
\end{eqnarray}
The integral in Eq. (\ref{NumintPol}) is of the form 
   $\int \mathrm{d} \theta  f(\theta)e^{ig(\theta)}$ and may be approximated through the method of the steepest descent \citep{Appel2007}. For $R/Fr^2>1$, the integrand oscillates rapidly and there are two stationary points given by $g'(\theta)=0$:
\begin{subeqnarray}
\theta_1(\phi)&=&  \frac12 (\arcsin (3\sin\phi)-\phi)  \ , \label{theta1}   \\
\theta_2(\phi)&=&         \frac12 (\pi -\arcsin (3\sin\phi)-\phi) \label{theta2}  \ .
\end{subeqnarray}
One shall note that at the Kelvin angle  $\phi = \phi_K=\arcsin(1/3)$, the two points $\theta_1$ and $\theta_2$ coalesce and thus the saddle-point method won't be accurate in the vicinity of $\phi = \phi_K$. The calculation for two coalescing saddle points \citep{Johnson1997} won't be developed it here as our aim is to study the behaviour of $\phi_{{max}}$ at large Froude numbers for which \textit{a priori} $\phi_{{max}}$ is far below $\phi_{ K}$. In this range both saddle points  can safely be considered independently.
Hence, far below $\phi_{ K}$ one can write:
\begin{eqnarray}
\breve Z(R,\phi)&\simeq& i\pi \left(\breve Z_1(R,\phi)+\breve Z_2(R,\phi)  \right)   \ ,    
\end{eqnarray}
where:
\begin{eqnarray}
\breve Z_{j}(R,\phi)\,\,=
  \sqrt{  \frac{2\pi}{\left|\partial_{\theta}^2g(R,\theta_{j},\phi)\right|}   } f(\theta_{j})   e^{i\left(g(R,\theta_{j},\phi)+\frac \pi4  \right)}  \, , \label{Z2} 
\end{eqnarray}
where $\theta_j$, $j\in \{1,2\}$,  are implicit function of $\phi$ as defined through Eq.~(\ref{theta2})  and where:
\begin{eqnarray}
f(\theta)&=&   \frac{ \widehat{P}(K_0(\theta),\theta)} {Fr^4\cos^4 \theta}\ , \\
g(R,\theta,\phi)&=&  - \frac{R \cos (\theta +\phi) }{Fr^2\cos^2\theta }\  . \label{g}
\end{eqnarray}
\smallskip
One can easily check that far below $\phi_{ K}$ the function $\breve  Z_1$ exclusively defines the transverse waves whereas $\breve  Z_2$ exclusively defines the diverging waves. Let us again take the Gaussian pressure field of Eq. (\ref{Gauss}). As an effect of the normalization of the pressure field, both functions decrease as the Froude number is increased. Yet, the amplitude of the $\breve Z_1$ waves scales as $ {Fr}^{-4}$  whereas the amplitude of the $\breve Z_2$ waves  scales as $Fr^{-3/2}$ thus decreasing slower. This explains why the transverse waves vanish as compared to the diverging waves for large Froude numbers.  Let us now thus focus on the $\breve  Z_2$ function.  At a given $R$, the angular envelope function of $\breve Z_2$ is given by: 
\begin{eqnarray}
h(R,\phi)&=&    \sqrt{  \frac{2\pi}{|\partial_{\theta}^2g(R,\theta_2,\phi)|}   }         f(\theta_2)    \ ,   \label{h}
\end{eqnarray} 
where $\theta_2$ and $g$ are defined through Eqs.~(\ref{theta2}) and (\ref{g}).   Figure 3 displays $\breve Z_2(R=10\Lambda,\phi)$ as given by Eq. (\ref{Z2})    as a function of $\phi$ for different Froude numbers. Their angular envelopes given by Eq. (\ref{h}) are signified with a solid black line.
\begin{figure}
\begin{center}
\includegraphics[scale=0.36]{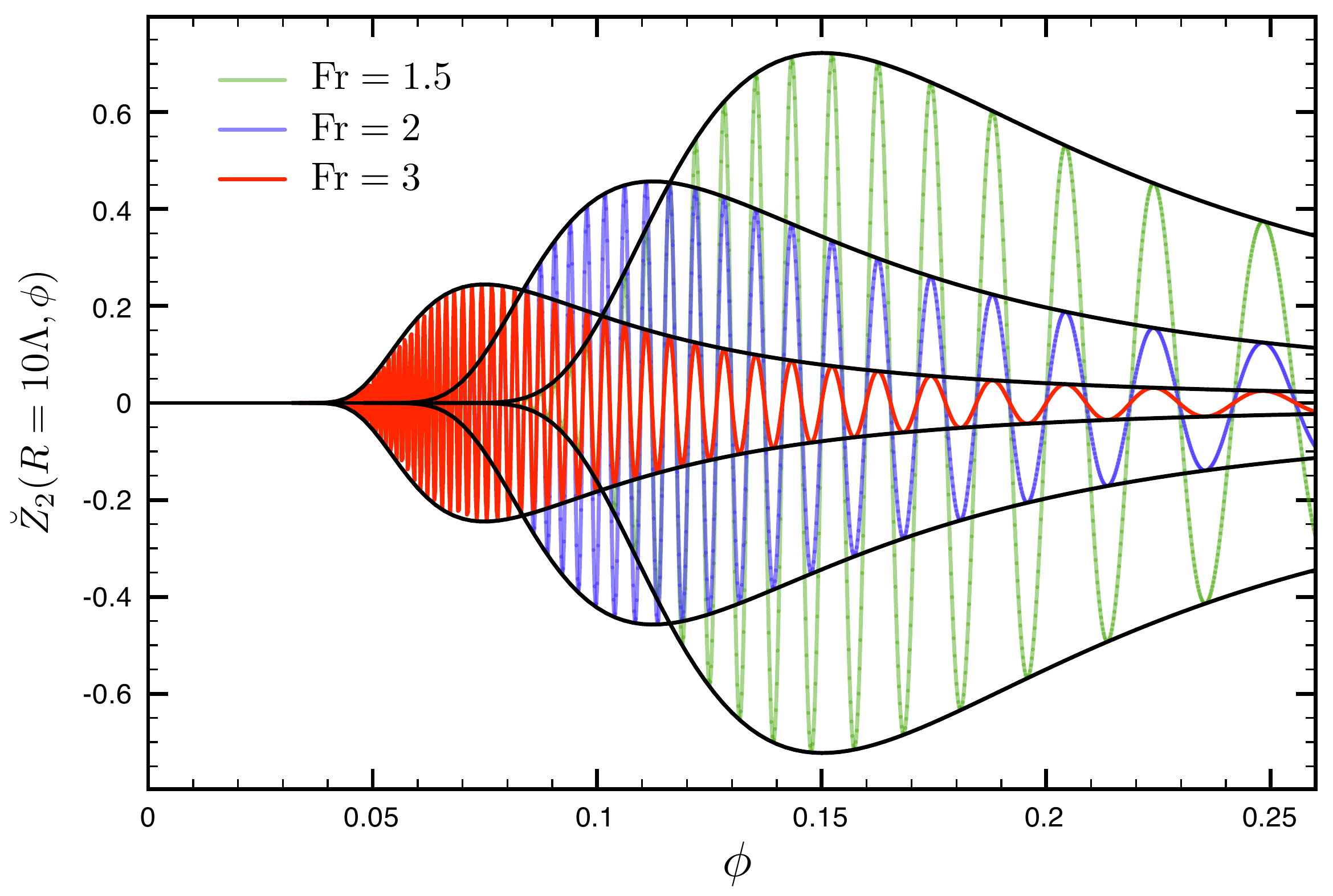}
\caption{Colour online. Plot of $\breve Z_2(R=10\Lambda,\phi)$ as given by Eq. (\ref{Z2})   with the Gaussian pressure field of Eq.~(\ref{Gauss}) where $\Lambda=2\pi Fr^2$ as a function of $\phi$ for different Froude numbers. Their angular envelopes given by Eq. (\ref{h}) are signified with a solid black line.}
\label{Arc}
\end{center}
\end{figure}
For small angles $\phi$, Eq. (\ref{h}) reduces to:
\begin{eqnarray}
h(R,\phi)&\simeq&  \sqrt{\frac{\pi}{16R}}\,\frac{1}{Fr^3\phi^{5/2}} \, \widehat{P}\left(\frac{1}{4 Fr^2 \phi^2} \, , \, \frac{\pi}{2} - 2\phi\right) \ . \label{hbis}
\end{eqnarray}
 The angle $\phi_{max}$ corresponding to the maximum of the amplitude is obtained by solving $\partial_\phi h=0$. In the case of an axisymmetric pressure field, this yields:
\begin{eqnarray}
5\widehat P(u)+4u\,\widehat P'(u)&=&0 \ , \label{Elie}
\end{eqnarray}
where $u=1/(2 Fr \,\phi)^2$. The solution $u^*$ of Eq.~(\ref{Elie}) being a pure number, the angle $\phi_{max}$ scales as: 
\begin{eqnarray}
\phi_{max}&\sim& \frac{1}{Fr} \ ,
\end{eqnarray}
since $u^*=1/(2 Fr\, \phi_{max})^2$.
In the particular case of the Gaussian pressure field introduced in Eq.~(\ref{Gauss}), one has:
\begin{eqnarray}
\phi_{{max}}  &=& \frac{1}{40^{1/4}  \sqrt{\pi}} \,\frac1{Fr}  \ . \label{scaling}
\end{eqnarray}
As one can see on Fig.~\ref{Beta}, this prediction (blue line) fits perfectly the numerical results at large Froude numbers.
\section{Conclusion}\label{sec:conclusion}
In this paper we performed a theoretical study of the Kelvin wake pattern generated by a moving perturbation and focused on the large Froude number regime. We showed that the angle delimiting the wake region outside which the surface is unperturbed remains constant and equal to the Kelvin angle for all Froude numbers. However, a different angle corresponding to the maximum of the amplitude of the waves can be identified. Considering an axisymmetric pressure field, we analytically showed that this angle  scales as $Fr^{-1}$ for large Froude numbers thus behaving as a Mach angle, as highlighted by Rabaud and Moisy in their observations of real ships.

\section{Acknowledgements}\label{sec:acknowledgements}
We wish to thank V. Bacot, O. Dauchot, S. Ellingsen, G. Falkovich, T. Salez and L. Tuckerman for very interesting discussions.

\bibliographystyle{jfm}

\bibliography{biblio}

\begin{thebibliography}{25}
\expandafter\ifx\csname natexlab\endcsname\relax\def\natexlab#1{#1}\fi

\bibitem[Appel(2007)]{Appel2007}
{\sc Appel, W.} 2007 {\em {Mathematics for Physics and Physicists}\/}.
  Princeton University Press.

\bibitem[Benzaquen {\em et~al.\/}(2011)Benzaquen, Chevy \&
  Rapha\"{e}l]{Benzaquen2011}
{\sc Benzaquen, M., Chevy, F. \& Rapha\"{e}l, E.} 2011 {Wave resistance for
  capillary gravity waves: Finite-size effects}. {\em Europhys. Lett.\/} {\bf
  {\bf96}}~(3), 34003.

\bibitem[Benzaquen \& Rapha\"{e}l(2012)]{Benzaquen2012}
{\sc Benzaquen, M. \& Rapha\"{e}l, E.} 2012 Capillary-gravity waves on
  depth-dependent currents: Consequences for the wave resistance. {\em
  Europhys. Lett.\/} {\bf {\bf97}}~(1), 14007.

\bibitem[Casling(1978)]{Casling1978}
{\sc Casling, E.~M.} 1978 Planing of a low-aspect-ratio flat ship at infinite
  froude number. {\em Journal of Engineering Mathematics\/} {\bf {\bf12}},
  43--57.

\bibitem[Chepelianskii {\em et~al.\/}(2008)Chepelianskii, Chevy \&
  Rapha\"{e}l]{Chepelianskii2008}
{\sc Chepelianskii, A.~D., Chevy, F. \& Rapha\"{e}l, E.} 2008
  {Capillary-Gravity Waves Generated by a Slow Moving Object}. {\em Phys. Rev.
  Lett.\/} {\bf {\bf100}}~(7), 074504.

\bibitem[Crawford(1984)]{Crawford1984}
{\sc Crawford, F.} 1984 Elementary derivation of the wake pattern of a boat.
  {\em Am. J. Phys.\/} {\bf {\bf52}}, 782.

\bibitem[Cumbertach(1958)]{Cumbertach1958}
{\sc Cumbertach, E.} 1958 Two-dimensional planing at high {Froude} number. {\em
  J. Fluid Mech.\/} {\bf {\bf4}}, 466--478.

\bibitem[Darrigol(2005)]{Darrigol2005}
{\sc Darrigol, O.} 2005 {\em {Words of Flow: A History of Hydrodynamics from
  the Bernoullis to Prandtl}\/}. Oxford University, New York.

\bibitem[Falkovich(2011)]{Falkovich2011}
{\sc Falkovich, G.} 2011 {\em {Fluid Mechanics - A Short Course for
  Physicists}\/}. Cambridge University Press.

\bibitem[Havelock(1908)]{Havelock1908}
{\sc Havelock, T.~H.} 1908 The propagation of groups of waves in dispersive
  media, with application to waves on water produced by a traveling
  disturbance. {\em Proc. R. Soc. A\/} {\bf {\bf95}}, 354.

\bibitem[Havelock(1919)]{Havelock1919}
{\sc Havelock, T.~H.} 1919 Periodic irrotational waves of finite height. {\em
  Proc. R. Soc. A\/} {\bf {\bf95}}, 38--51.

\bibitem[Johnson(1997)]{Johnson1997}
{\sc Johnson, R.~S.} 1997 {\em {A Modern Introduction to the Mathematical
  Theory of Water Waves}\/}. Cambridge University Press.

\bibitem[Kelvin(1887)]{Kelvin1887}
{\sc Kelvin, Lord} 1887 On the waves produced by a single impulse in water of
  any depth. {\em Proc. R. Soc. London, Ser. A\/} {\bf {\bf42}}, 80--83.

\bibitem[Lai \& Troesch(1995)]{Lai1995}
{\sc Lai, C. \& Troesch, A.W.} 1995 Modelling issues related to the
  hydrodynamics of three-dimensional steady planing. {\em Journal of Ship
  Research\/} {\bf {\bf39}}, 1--24.

\bibitem[Lamb(1993)]{Lamb1993}
{\sc Lamb, H.} 1993 {\em {Hydrodynamics}\/}. Cambridge University Press,
  Cambridge, 6th ed.

\bibitem[Le~Merrer {\em et~al.\/}(2011)Le~Merrer, Clanet, Qu\'er\'e,
  Rapha\"{e}l \& Chevy]{LeMerrer2011}
{\sc Le~Merrer, M., Clanet, C., Qu\'er\'e, D., Rapha\"{e}l, E. \& Chevy, F.}
  2011 Wave drag on floating bodies. {\em Proc. Natl. Acad. Sci\/} {\bf
  {\bf108}(30)}, 15064Ð15068.

\bibitem[Lighthill(1978)]{Lighthill1978}
{\sc Lighthill, J.} 1978 {\em {Waves in Fluids}\/}. Cambridge University Press,
  Cambridge.

\bibitem[Nakos \& Sclavounos(1990)]{Nakos1990}
{\sc Nakos, D.~E. \& Sclavounos, P.~D.} 1990 On steady and unsteady ship wave
  pattern. {\em J. Fluid Mech.\/} {\bf {\bf215}}, 263--288.

\bibitem[Parnell \& Kofoed-Hansen(2001)]{Parnell2001}
{\sc Parnell, K.~E. \& Kofoed-Hansen, H.} 2001 Wakes from large high-speed
  ferries in confined coastal waters: Management approaches with examples from
  {New-Zealand and Denmark}. {\em Coastal Management\/} {\bf {\bf29}},
  217--237.

\bibitem[Rabaud \& Moisy(2013)]{rabaud2013ship}
{\sc Rabaud, M. \& Moisy, F.} 2013 Ship wakes: Kelvin or mach angle? {\em Phys.
  Rev. Lett\/} {\bf {\bf110}}, 214503.

\bibitem[Rabaud \& Moisy(26-28 June 2013)]{Rabaud2013Autre}
{\sc Rabaud, M. \& Moisy, F.} 26-28 June 2013 Narrow ship wakes and wave drag
  for planning hulls. {\em Innov-Sail, Lorient\/} .

\bibitem[Rapha{\"e}l \& de~Gennes(1996)]{Raphael1996}
{\sc Rapha{\"e}l, E. \& de~Gennes, P.-G.} 1996 Capillary gravity waves caused
  by a moving disturbance: wave resistance. {\em Phys. Rev. E\/} {\bf
  {\bf53}}~(4), 3448.

\bibitem[Suzuki {\em et~al.\/}(21-23 July 1997)Suzuki, Nakata, Ikehata \&
  Kai]{Suzuki1997}
{\sc Suzuki, K., Nakata, Y., Ikehata, M. \& Kai, H.} 21-23 July 1997 Numerical
  prediction on wave making resistance of high speed trimaran. {\em Fourth Int.
  Conf. on Fast Sea Transportation, Sydney\/} .

\bibitem[Tuck {\em et~al.\/}(8-13 July 2002)Tuck, Scullen \&
  Lazauskas]{Tuck2002}
{\sc Tuck, E.~O., Scullen, D.~C. \& Lazauskas, L.} 8-13 July 2002 Wave patterns
  and minimum wave resistance for high-speed vessels. {\em 24th Symposium on
  Naval Hydrodynamics, Fukuoka, Japan\/} .

\bibitem[Voise \& Casas(2010)]{Voise2010}
{\sc Voise, J. \& Casas, J.} 2010 The management of fluid and wave resistances
  by whirligig beetles. {\em J. R. Soc. Interface\/} {\bf {\bf7}}, 343.

\end{thebibliography}

\end{document}